\documentstyle[aps,aipbook,floats]{revtex}

\newcommand{\be}{\begin{equation}}
\newcommand{\bea}{\begin{eqnarray}}
\newcommand{\ba}{\begin{array}}
\newcommand{\bean}{\begin{eqnarray*}}
\newcommand{\ee}{\end{equation}}
\newcommand{\eea}{\end{eqnarray}}
\newcommand{\ea}{\end{array}}
\newcommand{\eean}{\end{eqnarray*}}
\newcommand{\hPi}{\mbox{$\widehat{\Pi}$}}
\newcommand{\hGa}{\mbox{$\widehat{\Gamma}$}}

\newcommand{\gi}{gauge-invariant}

\def \dsl {\partial \kern-.55em{/}}
\def \Dsl {D \kern-.65em{/}}
\def \qsl {q \kern-.45em{/}}
\def \slp {p \kern-.45em{/}}
\def \ksl {k \kern-.45em{/}}
\def \0sl {0 \kern-.65em{/}}

\def \Gfs {Green's functions~}

\def \per {perturbation~}

\def \gd  {gauge dependence~}
\def \gdt  {gauge dependent~}
\def \git  {g.i.~}
\def \gi  {gauge independence~}
\def \S  {S-matrix~}
\def \g5 {\gamma _5}

\begin{document}
\title{The Ward Identities of the Gauge \\ Invariant  Three Boson Vertices}
\author{Kostas Philippides}
\address{ New York University,
 New York, New York 10003}
\maketitle
\begin{abstract}
We outline the pinch technique for constructing gauge invariant
\Gfs in  gauge theories and
derive the Ward identities that must be satisfied by the
gauge invariant three boson  vertices of the standard model.
They are  generalizations of their  tree level  Ward identities
and are shown to be crucial for the delicate gauge cancellations
of the \S.
\footnote{To appear in the proceedings of the International Symposium on
Vector Boson Self Interactions, Feb. 1-3, 1995  UCLA.}
\end{abstract}

\section*{The Pinch Technique}
As is well known,
the {\it off-shell~}     Green's functions
of non Abelian gauge theories
are in general
gauge dependent as a result of the quantization procedure.
For example in the t'Hooft covariant
class of gauges
$R_{\xi}$, the presence of the gauge fixing
term and the Fadeev--Popov  ghost term
give rise to
gauge boson, ghost, and unphysical scalar
propagators that depend explicitly on the unphysical gauge parameters
$\xi_i ,~ i= g, \gamma, Z, W$.
The gauge boson propagators are given by
\be
\Delta^{\mu\nu}_{i}(q,\xi_i)= \frac{1}{q^{2}-M^{2}_{i}}
[g^{\mu\nu} - (1-\xi_{i})\frac{q^{\mu}q^{\nu}}
{q^{2}-\xi_{i}M^{2}_{i}}]~~,
\label{Prop}
\ee
and  the \Gfs of the theory are infected by  the
unphysical $\xi_i$ parameters
through the bosonic quantum loop corrections.

Although  gauge invariance has been lost upon quantization, the effective
quantum action still exhibits residual symmetries that lead to  Ward
identities (WI) between the \gdt \Gfs. In particular the BRST symmetry
gives rise
to complicated Slavnov-Taylor (ST)  identities that make explicit reference to
the gauge parameters and involve the ghosts' \Gfs
while the background field gauge symmetry
preserves the tree level WI
but only for the  background \Gfs$\!$.

In spite of the fact that the \Gfs  of the theory are \gdt,
the S-matrix  is gauge independent (g.i.)
order by order in \per  theory,
 as a result of a subtle
gauge cancellation between self--energy, vertex, and box graphs.
Nevertheless the \gd
of the \Gfs is considered as an unhappy state of affairs for at least two
reasons: i) In non--perturbative studies
the infinite set of the SD equations is built out of the \gdt
Green's functions; when
casually truncated residual gauge dependences infest ostensibly \git
quantities giving rise to meaningless approximations.
ii) In perturbative studies form factors
extracted from the usual vertices 
are gauge dependent, and thus void of  any
physical meaning, 
which makes the generalization
of the familiar classical moments problematic at the quantum level.

The  pinch technique, originally proposed by Cornwall \cite{Corn1},
\cite{Corn2},  has successfully addressed these issues.
It exploits  all the healthy properties of the \S
and allows the construction
of modified \git $n$-point functions, through the
order by order rearrangement of
Feynman graphs, contributing to a certain physical
and therefore 
 g.i. amplitude.

The \Gfs constructed via the PT satisfy the following properties:
i)
They are gauge parameter independent and ii) gauge fixing procedure
independent as has been shown by explicit calculations in a wide variety
of gauges ($R_{\xi}$, Unitary, Axial, $R_{\xi_Q}$)
\cite{Corn2},\cite{DeSi},\cite{PaSi},\cite{PaBFM}.
ii) They are process independent \cite{Watson}
iii) Running couplings defined directly from  the PT self energies $\hPi$,
obey the RG equations \cite{Corn2},\cite{DeSi}, and  the form
factors extracted from the PT vertices $\hGa$,  are well behaved \cite{Pa90},
\cite{PaPhi1}
iv) Finally  they satisfy naive QED-like WI which are just their
respective  tree level
classical WI \cite{CoPa},\cite{Pa90},
\cite{PaPhi1}, \cite{PaSe}, \cite{PaPhi2}.

Although in principle the PT can be applied to any order,
so far explicit results have been obtained
only up to one loop.


\section*{The Ward Identities}

After the PT rearrangement
has been completed,
the amplitude we consider has been reorganized
into individually \git loop  structures (Green's functions)
connected by  \gdt tree level propagators.
In other words, the PT algorithm only cancels
all gauge dependences originating from
the tree-level propagators appearing {\it inside} the loops, but
a residual gauge dependence, stemming from
boson propagators {\it outside} of loops, survives at the end of the
pinching process.

When the external currents are conserved any residual \gd will
automatically cancel in the $R_{\xi}$ gauges.
However this is not the case in the axial gauges
and when the currents are not conserved this remaining \gd will persist
even in the $R_{\xi}$ gauges.


The cancellation of this remaining  \gd from the \S  becomes possible
due to a set of WI satisfied by the \git \Gfs.
One can actually derive these WI {\it without}
any detailed knowledge of the algorithm which gives rise to the \git
Green's functions.
All one needs to assume
is that such an algorithm exists
(in our case the PT algorithm), and
demand  the complete
\gi of the \S$\!$.
So, once
the \git \Gfs have been constructed,
one should examine whether or not they
actually satisfy the
required WI, as a self-consistency check.

It is instructive
to illustrate the derivation of the  WI for the simple case of the
g.i. $W$ propagator. We consider the one-loop
S-matrix element of the charged current process
$ e^- + \nu_e \rightarrow \nu_e +  e^-  $
and apply the PT rules in the context of the $R_{\xi}$ gauges.
After the application of the PT,
the part of the S-matrix ~$\widehat{T}_1(q^2)$, which only depends
on the momentum transfer  $q^2$, will consist of four amplitudes each one
corresponding to the relevant \git 2-point function
 of the $W$ and its associated
unphysical scalar $\phi$.
We use the  current relation
$q_{\mu} J^{\mu}_{W^{\pm}} = iM_{W} J_{\phi^{\pm}}$
in order to
pull out a common factor
$q_{\alpha}J^{\alpha}_{W^{+}} q_{\beta}J^{\beta}_{W^{-}}$ from  all these
 amplitudes  and  then employ the standard identity
\be
\Delta^{\mu\nu}_{W}(q,\xi_W)=
U^{\mu\nu}_{W}(q)-\frac{q^{\mu}q^{\nu}}{M^{2}_{W}}\Delta_{\phi}(q,\xi_W)~~,
\label {Id1}
\ee
in order to isolate all remaining gauge dependences into scalar
propagators  of the $\phi$, which is given by
\be
\Delta_{\phi}(q,\xi_W) = \frac{-1}{q^2-\xi_W M_W^2}~.
\ee
$U^{\mu\nu}_{W}(q)$ is the
\git propagator of the $W$ at tree level, namely the propagator in the
unitary gauge.
After this last step one observes that the remaining gauge
dependences will arise from terms containing either one or
two $\Delta_{\phi}$.
The requirement that ~${\hat{T}}_{1}(q^2)$ is
$\xi_W$-independent, gives rise to
two independent equations; the first enforces
the cancellation of the terms
with only one $\Delta_{\phi}$ factor, whereas the second
enforces the cancellation of the terms
with a $\Delta_{\phi}~\Delta_{\phi}$ factor.
It then follows that
\be
q^{\mu}\hPi^W_{\mu\nu}(q) \mp iM_W \hPi^{\pm}_{\nu}(q) = 0
\label{W1}
\ee
\be
q^{\mu}\hPi^{\pm}_{\mu}(q) \pm iM_W \hPi^{\phi}(q) = 0
\label{W2}
\ee
\be
q^{\mu}q^{\nu}\hPi^W_{\mu\nu}(q) - M^2_W \hPi^{\phi}(q) = 0
\label{W3}
\ee
Along the same line of reasoning similar
 WI can be obtained for the Z and $\chi$ self energies. For the photon
and the gluon, the relevant WI are obtained by repeating the above
arguments in an axial gauge.

We now turn to our main
ob
$\!$$\!$jective, namely the derivation of the WI for
the g.i three boson vertices (TBV). In order to construct a
g.i. $n$-point function in the context of the \S PT one has to employ
an $n \rightarrow n$ procces. We choose to use fermions as external test
particles and the process
$
e^- (n) + \nu(\ell)  +  e^-(r) \to
e^-(\hat{n}) + e^-(\hat{\ell})
+  \nu (\hat{r})
$
where
$
q = n - \hat{n} ~,~
p_{1}~ = \ell - \hat{\ell} ~,~
 p_{2}~ = r - \hat{r} ~,~
$
are the momentum transfers at the corresponding fermion lines;
they represent the incoming momenta of each of the bosons,
merging in the TBV. Following  the PT prescription we identify the
pinch parts and allot  them to the relevant graphs ending up
 with
g.i. ``blobs'' (\Gfs$\!$)
connected by \gdt tree level bosonic propagators. Concentrating
on the amplitudes from which an overall factor containing the three external
fermionic currents can be pulled out, we use  the identity of
Eq.(\ref{Id1}) and its $Z,\chi$ analogue to isolate all remaining
gauge dependences in the form of propagators of the unphysical scalars.
In this case there are many gauge dependent terms displaying
characteristic structures that depend on the kind and
number of scalar  propagators they contain, and
the momenta carried by these propagators. Clearly if two such terms differ in
any of the above three aspects they are linearly independent and the cofactor
in front of them must vanish individually. We are thus led to a number of
conditions, which enforce the remaining  gauge cancellations
and give rise to a tower of WI
satisfied by the three boson vertices and relating them to
the boson self-energies. In
particular the final cancellation of the $\Delta_{\chi}(q,\xi_Z)$ terms
leads to
\be
q^\mu \hGa ^{ZW^-W^+} _{\mu \alpha \beta}
+ iM_Z \hGa ^{\chi W^-W^+} _{\alpha \beta}
 =  gc ~\left[ \hPi ^W _{\alpha \beta} (p_1) - \hPi ^W _{\alpha \beta} (p_2)
\right]~,
\label{qZWW}
\ee
while the  cancellation of the $\Delta_{\phi}(p_1,\xi_W)$ and
$\Delta_{\phi}(p_2,\xi_W)$ requires respectively
\be
p_{1}^\alpha \hGa ^{V W^-W^+} _{\mu \alpha \beta}
+i M_W \hGa ^{V\phi^- W^+} _{\mu \beta}
 =  g_V  \left[ \hPi ^W _{\mu \beta} (p_2)
                  - \hPi ^{V} _{\mu \beta} (q)
                - \frac {g_{V'}} {g_V} \hPi ^{V V'} _{\mu \beta} (q)
  \right]~,
\label{p1gWW}
\ee
\be
\!p_{2}^\beta \hGa ^{VW^-W^+} _{\mu \alpha \beta}
+iM_W \hGa ^{VW^-\phi^+}  _{\mu \alpha}
 = g_V \left[ \hPi ^{V} _{\mu \alpha} (q)
              + \frac{g_{V'}} {g_V} \hPi ^{V V'} _{\mu \alpha} (q)
                 ~- \hPi ^W _{\mu \alpha} (p_1)\right]~,
\label{p2gWW}
\ee
where $V=\gamma,Z$, $g_{\gamma} = gs$, $g_{Z} = gc$, and $s^2=1-c^2$ is an
abbreviation for $\sin^2\theta_W$.
To derive the analogue of Eq.(\ref{qZWW}) for the
$\hGa^{\gamma W^-W^+} _{\mu \alpha \beta}$
vertex  we have to work in an axial gauge,
in which case we obtain
\be
 q^\mu \hGa ^{\gamma W^-W^+} _{\mu \alpha \beta}
 =  gs \left[ \hPi ^W _{\alpha \beta} (p_1) - \hPi ^W _{\alpha \beta} (p_2)
\right]~.
\label{qgWW}
\ee
Finally WI where the g.i. TBV are contracted with two or three momenta
can be easily derived, by demanding the cancellation of gauge dependences
stemming from the terms with more than one unphysical scalar propagators
e.g.  $\Delta_{\chi}(q,\xi_Z)\Delta_{\phi}(p_1,\xi_W)$ etc. .

All of the above WI for the PT \Gfs have been explicitly proven up to
one loop order, for some cases in  more than one gauge. We emphasize that they
constitute an extension, to higher orders, of the tree level WI stemming
from the classical gauge invariance, and in this respect they are QED-like.
These same  WI  also allow  a  consistent g.i. truncation
procedure for the SD equations of the theory.
In contrast to the ST identities,  they are simple, manifestly g.i. and make
no reference to ghost \Gfs. Since they are just the tree level WI,
 they coincide
with the WI satisfied by the background \Gfs
of the background field method (BFM) but
contrary to the latter they are not restricted to background \Gfs only,
they relate g.i. quantities and
can be obtained within any gauge fixing procedure.

The  derivation of the WI presented above, displays
in a very transparent way the mechanism responsible for the
 gauge cancellations of the S-matrix.
The amplitudes of an \S reorganize
themselves through
pinching; this suffices to cancel all gauge dependences inside loops and
makes possible the definition of g.i. \Gfs$\!$. The \Gfs thus constructed
satisfy their tree level WI,  which enforce the elimination of
all remaining gauge dependences that were initially present
outside of the loops. This implies
that, the gauge dependences originating inside the loops cancel
independently
from the ones residing  outside of loops,
which points towards an interesting property of the \S  already known to
hold true in QED. Namely,
one could freely choose  {\it two different}
$\xi$-parameters $\xi_l$ and $\xi_t$,
to gauge fix the bosonic propagators that appear inside and outside
of the loops respectively.
Although this dual choice of gauges
 seems quite  arbitrary and lacks any field theoretical
justification at a formal level, the WI guarantee that the \S will still be
gauge invariant and independent of both $\xi_t$ and  $\xi_l$.

In closing, we point out that
the WI of Eqs.(\ref{qZWW},\ref{qgWW}) can find  immediate application,
in incorporating,  the full  one loop $W$-width effects
in the tree level cross sections of
processes involving the production or decay of on-shell $W$s,
in a way consistent with gauge invariance.

\section*{Acknowledgments}

I would like to thank J.~M.~Cornwall, and J.~Papavassiliou for useful
discussions, and  E.~Karagiannis for his kind hospitality during my
visit in Los Angeles.


\end{document}